\RequirePackage{ifpdf}
\documentclass{PoS}
\pdfoutput=1

\usepackage{subfig}
\usepackage{float}
\usepackage{url}
\usepackage{braket}
\usepackage{amsmath}
\usepackage{amssymb}
\usepackage{esint}
\usepackage{dsfont}
\usepackage{eurosym}
\usepackage[utf8]{inputenc}
\usepackage[T1]{fontenc}
\usepackage{verbatim}

\usepackage{array}
\usepackage{tabularx}
\usepackage{booktabs}
\usepackage{rotating}
\usepackage{algorithm}
\usepackage{algorithmic}

\graphicspath{{figures00/}}
\graphicspath{{figures/}}

\newcommand{\Tr}{\mbox{\rm Tr\,}}
\newcommand{\Real}{\mbox{\rm Re\,}}

\usepackage{cleveref}
\crefname{equation}{equation}{equations}
\Crefname{equation}{Equation}{Equations}
\crefname{table}{table}{tables}
\Crefname{table}{Table}{Tables}
\crefname{figure}{figure}{figures}
\Crefname{figure}{Figure}{Figures}
\crefname{algorithm}{algorithm}{algorithms}
\Crefname{algorithm}{Algorithm}{Algorithms}

\title{Coulomb and Landau Gauge Fixing in GPUs using CUDA and MILC}

\ShortTitle{Coulomb and Landau Gauge Fixing in GPUs using CUDA and MILC}

\author{\speaker{Nuno Cardoso} \\
NCSA, University of Illinois\\
E-mail: \email{ncardoso@illinois.edu}
}

\abstract{
    In this work, we present the GPU implementation of the overrelaxation and steepest descent method with Fourier acceleration methods for Laudau and Coulomb gauge fixing using CUDA for SU(N) with N>2. 
A multi-GPU implementation of the overrelaxation method is also presented using MPI and CUDA. 
The GPU performance was measured on BlueWaters and compared against the gauge fixing of the CPU MILC code.
}

\FullConference{The 32nd International Symposium on Lattice Field Theory\\
                 23-28 June, 2014\\
                 Columbia University New York, NY}

\begin{document}

\section{Introduction}

On the lattice, the Coulomb/Landau gauge is defined by maximising the functional
\begin{equation}
	F_U[g]=\frac{1}{4 N_cV}\sum_x\sum_\mu\Real\left[ \Tr\left(  g(x)U_\mu(x)g^\dagger(x+\hat{\mu}) \right) \right]
\end{equation}
with $N_c$ the dimension of the gauge group and $V$ the lattice volume.
On the gauge fixing process, the quality of the gauge fixing is measured by
\begin{equation}
	\theta = \frac{1}{N_c V}\sum_x \Tr\left[\Delta(x)\Delta^\dagger(x)\right]
\end{equation}
where
\begin{equation}
	\Delta(x) = \sum_\nu\left[ U_\nu(x-a\hat{\nu})-U_\nu(x) - \text{h.c.} - \text{trace} \right]
\end{equation}
is the lattice version of $\partial_\mu A_\mu=0$, where $\mu=0,1,2,3$ for landau gauge and $\mu=0,1,2$ for the Coulomb gauge (where the temporal direction is along $\mu=3$).	
Two well known methods to fix the gauge are the relaxation algorithm via overrelaxation and the steepest descent method with FFTs.

The relaxation algorithm aims to optimize the value of $F_U[g]$ locally, i.e., searching the maximum of
\begin{equation}
	f^g(x)=\Real\Tr\left[g(x)K(x)\right]
\end{equation}
for all $x$, where
\begin{equation}
	K(x)=\sum_\mu \left(U_\mu(x)g^\dagger(x+\hat{\mu}) + U_\mu^\dagger(x-\hat{\mu})g^\dagger(x-\hat{\mu})\right)
\end{equation}
The local solution is then given by
\begin{equation}
	g(x) = K^\dagger(x)\left(\det K^\dagger(x)\right)^{-1/2}
\end{equation}
in the case of the gauge group SU(2).
For $N > 2$ one iteratively operates in the $(N(N-1)/2)$ SU(2) subgroups.
The overrelaxation algorithm, \cite{Mandula:1990vs}, replaces the gauge transformation $g(x)$ by $g^\omega(x)$ with $\omega\in[1,2[$ in each step of the iteration.
The gauge fixing with overrelaxation algorithm is described in \cref{alg:Overrelaxation}.

The naive steepest descent method chooses at each step of the iterative procedure
\begin{equation}
	g(x) = \exp\left[\frac{\alpha}{2}\left( \sum_\nu \Delta_{-\nu}\left[U_\nu(x)-U_\nu^\dagger(x)\right]   -\text{trace} \right)\right]
\end{equation}
However, when it is applied to larger lattices, this method faces the problem of critical slowing down.	
This problem can be attenuated by Fourier acceleration\cite{Davies:1987vs}.
At each iteration one chooses
\begin{equation}
	g(x) = \exp\left[ \hat{F}^{-1} \frac{\alpha}{2}\frac{p^2_\text{max}a^2}{p^2a^2} \hat{F} \left( \sum_\nu \Delta_{-\nu}\left[U_\nu(x)-U_\nu^\dagger(x)\right]   -\text{trace} \right)\right]
\end{equation}
	with
\begin{equation}
	\Delta_{-\nu}\left(U_\mu(x)\right) = U_\mu(x-a\hat{\nu})-U_\mu(x)
\end{equation}
$p^2$ are the eigenvalues of $\left(-\partial^2\right)$, $a$ is the lattice spacing and $\hat{F}$ represents a fast Fourier transform (FFT). 
For the parameter $\alpha$, the optimal value is 0.08.
For numerical purposes, it is enough to expand to first order the exponential, followed by a reunitarization.
The gauge fixing algorithm using the steepest descent method with FFTs is described in \cref{alg:SteepestDescentFFTs}.

\noindent\begin{minipage}[T]{\textwidth}
  \begin{minipage}{.48\textwidth}
\begin{algorithm}[H]
\begin{algorithmic}[2] 
\STATE calculate $F_g[U]$ and $\theta$ (optional)
\WHILE{$\theta \geq \epsilon$}
\FOR{$\text{site parity} = $ even, odd } 
\FORALL{$x$ with same parity} 
\FORALL{SU(2) subgroups} 
\STATE{local optimization, find $g(x)\in SU(2)$
\STATE  which is function of $U_\mu(x)$ and  $U_\mu(x-\hat{\mu})$ } 
\FORALL{$\mu$} 
\STATE{apply $g(x)$ to $U_\mu(x)$ and  $U_\mu(x-\hat{\mu})$} 
\ENDFOR
\ENDFOR 
\ENDFOR
\ENDFOR
\STATE calculate $F_g[U]$ and $\theta$
\ENDWHILE
\end{algorithmic}
\caption{Overrelaxation algorithm.}
\label{alg:Overrelaxation}
\end{algorithm}
  \end{minipage}
  \begin{minipage}{.05\textwidth}
  \quad
  \end{minipage}
  \begin{minipage}{.48\textwidth}
\begin{algorithm}[H]
\begin{algorithmic}[2] 
\STATE calculate $\Delta(x)$, $F_g[U]$ and $\theta$
\WHILE{$\theta \geq \epsilon$}
\FORALL{ elements of $\Delta(x)$ matrix }
\STATE apply FFT forward
\STATE apply $p^2_\text{max}/p^2$
\STATE apply FFT backward
\STATE normalize 
\ENDFOR
\FORALL{$x$}
\STATE obtain $g(x)$ from $\Delta(x)$ and reunitarize
\ENDFOR
\FORALL{$x$}
\FORALL{$\mu$}
\STATE $U_\mu(x) \rightarrow g(x)U_\mu(x)g^\dagger(x+\hat{\mu})$
\ENDFOR
\ENDFOR
\STATE calculate $\Delta(x)$, $F_g[U]$ and $\theta$
\ENDWHILE
\end{algorithmic}
\caption{Steepest descent method with FFTs.}
\label{alg:SteepestDescentFFTs}
\end{algorithm}
  \end{minipage}
\end{minipage}

\section{Implementation}


The gauge links are stored in a 1-dimensional array with size $volume \times 4 \times Elems$ in the GPU global memory, with $Elems=4,6,9$ complex elements in the case of SU(3) and $volume=nx*ny*nz*nt$. The memory access is done as\\
$id = (i+j\times nx+k\times nx\times ny+t\times nx\times ny\times nz) / 2 + site\_parity \times volume / 2 + \mu \times volume$\\
where the even and odd sites are kept separate and each SU(N) complex element is stored with stride $4 \times volume$.

In overrelaxation method, each gauge link, $U_\mu(x)$, to be updated are dependent on the neighbors at $U_\mu(x)$ and $U_\mu(x-\hat{\mu})$, i.e., in total eight gauge links must be loaded from memory. 
Therefore, assigning one thread per lattice site leads to a high memory traffic and local memory usage.
A solution to avoid this is using eight threads per single lattice site, \cite{Schrock:2012fj}. 
Since the GPU warp size is 32, the minimum block size should be $32\times 8$ threads in order to maintain the memory reads coalesced from global memory and the grid size must be $volume / 32$.
The gauge links are loaded into registers whenever possible and the data exchange between threads is done through shared memory.
The $\sum_\mu$ can be done using only shared memory (more shared memory consumption) or using CUDA Atomic functions in shared memory, although double precision atomic additions are not yet supported natively).
The multi-GPU implementation was done using MPI. The code was implemented in order to support any type of lattice partition however all local lattice dimensions must be an even number.
At each gauge fixing step, the updates are first done in even lattice sites and then in the odd sites. Each node needs to exchange the top links with the same direction as the partitioned dimension and also since the exchanged links are updated we need to exchange them back.
To overlap the gauge link updates with node communication, we update first the bottom and top links in the partitioned dimensions and then overlap the exchange top links (current parity) and the ghost links (opposite parity) with the update of the remaining gauge links.
In order to support any kind of lattice partition, we pre-calculate the border and the interior lattice sites separately.
 
The steepest descent method with FFTs implemented here is a generalization of the implementation described in \cite{Cardoso:2012pv}. Here, we generalized the previous implementation to support even/odd lattice array and SU(N) with N>2.
Due to the high number of FFTs per gauge fixing step, $N\times N \times\left( \text{4D FFT} + \text{4D IFFT})\right)$ (in SU(3), using the 12 real number parametrization, instead of 9, we can reduce this to 6) and due to a lack of support for 4D FFTs in GPUs, we haven't done a multi-GPU implementation. 
However, one solution to avoid using FFTs is to use a multigrid implementation of the Fourier acceleration method \cite{Cucchieri:1998ew}.

\begin{figure}[h]
\begin{centering}
\subfloat{
\includegraphics[trim = 122mm 55mm 0mm 25mm, clip=true, height=0.65cm]{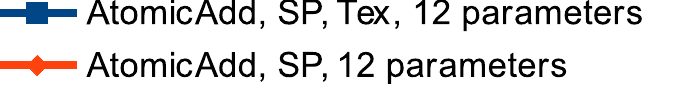}
\includegraphics[trim = 122mm 45mm 0mm 35mm, clip=true, height=0.65cm]{figures/bluew_landau_ovr_conv_gflops_preid_changeMILCorder}
\includegraphics[trim = 122mm 35mm 23mm 45mm, clip=true, height=0.65cm]{figures/bluew_landau_ovr_conv_gflops_preid_changeMILCorder}
\includegraphics[trim = 122mm 25mm 23mm 55mm, clip=true, height=0.65cm]{figures/bluew_landau_ovr_conv_gflops_preid_changeMILCorder}}
\setcounter{subfigure}{0}

\vspace{-0.3cm}\subfloat[Single GPU performance. \label{subfig:landau_single_ovr}]{
\begin{centering}
\includegraphics[trim = 6mm 0mm 78mm 9mm, clip=true, height=5.5cm]{figures/bluew_landau_ovr_conv_gflops_preid_changeMILCorder}
\par\end{centering}}
\subfloat[Multi-GPU performance, $32^4$  lattice volume.\label{subfig:landau_multi_ovr}]{
\begin{centering}
\includegraphics[trim = 6mm 0mm 57mm 9mm, clip=true, height=5.5cm]{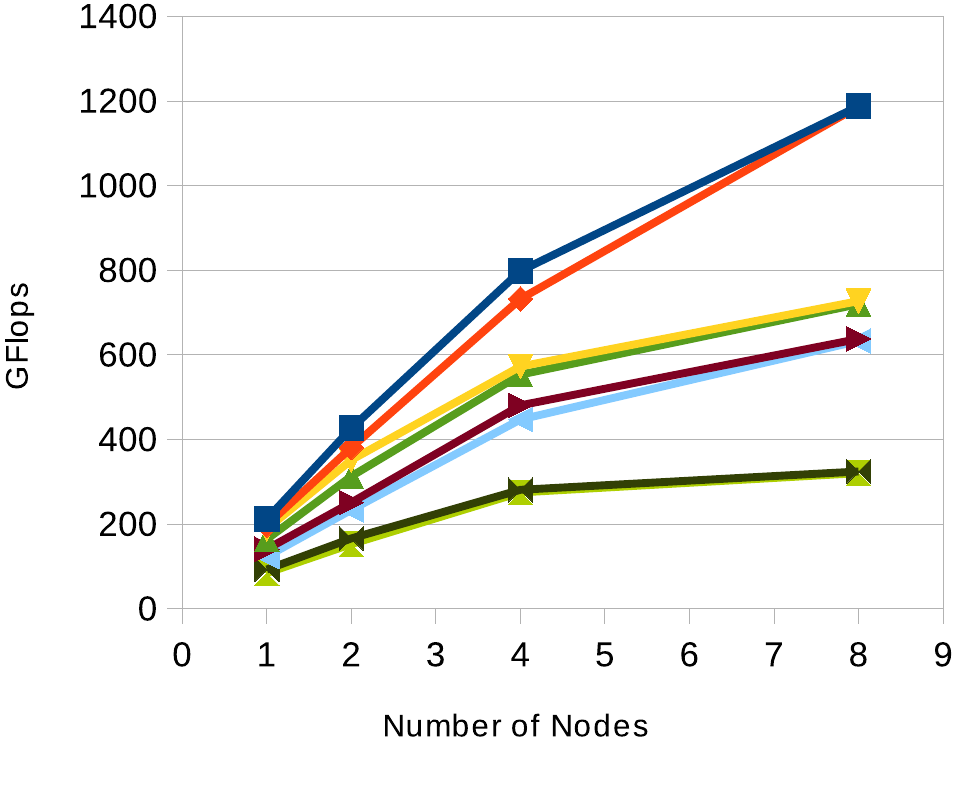}
\par\end{centering}}

\par\end{centering}
    \caption{SU(3) performance of the Overrelaxation algorithm. AtomicAdd means using CUDA atomicAdd function, SP/DP means single/double precision, Tex means using Texture memory, 12/18 parameters are the number of SU(3) real parameters used to store the gauge array in memory.}
    \label{fig:perf_overrelaxation_landau}
\end{figure}

\section{Results}

The performance tests were done in BlueWaters system with MILC, \cite{MILC}, and CUDA 5.5. Each node in BlueWaters system has one Tesla K20X GPU, with ECC code enabled by default.
In this paper, we only present the performance results for the SU(3) case. 
Despite the code for SU(3) supports global memory storage with 18 (full SU(3) matrix), 12 and 8 real parameters, here we only present the performance results for the full matrix and 12 real number parameterization.
We measured the performance for 1000 iterations with link reunitarization at each 20 steps. The overrelaxation boost parameter was set at 1.5 and in stepeest descent method with FFTs $\alpha=0.08$.

The MILC code by default splits the lattice between nodes by the following order $X\rightarrow Y\rightarrow Z\rightarrow T$, where $X$ is the lattice fastest index. 
Therefore, we changed the MILC code to allow the layout lattice partition by the following priority order $T\rightarrow Z\rightarrow Y\rightarrow X$ which is preferable for GPUs.

\begin{figure}[h]
\begin{centering}
\subfloat{
\includegraphics[trim = 0mm 32mm 0mm 0mm, clip=true, height=0.7cm]{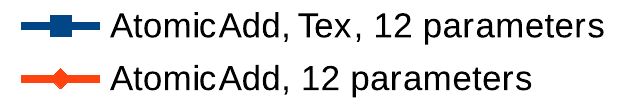}
\includegraphics[trim = 0mm 22mm 0mm 10mm, clip=true, height=0.7cm]{figures/BW_landau_weakscaling_legend-cropped}
\includegraphics[trim = 0mm 10mm 20mm 22mm, clip=true, height=0.7cm]{figures/BW_landau_weakscaling_legend-cropped}
\includegraphics[trim = 0mm 0mm 20mm 32mm, clip=true, height=0.7cm]{figures/BW_landau_weakscaling_legend-cropped}}
\setcounter{subfigure}{0}

\vspace{-0.3cm}\subfloat[Single precision.]{
\begin{centering}
\includegraphics[trim = 6mm 5mm 64mm 9mm, clip=true, height=5.5cm]{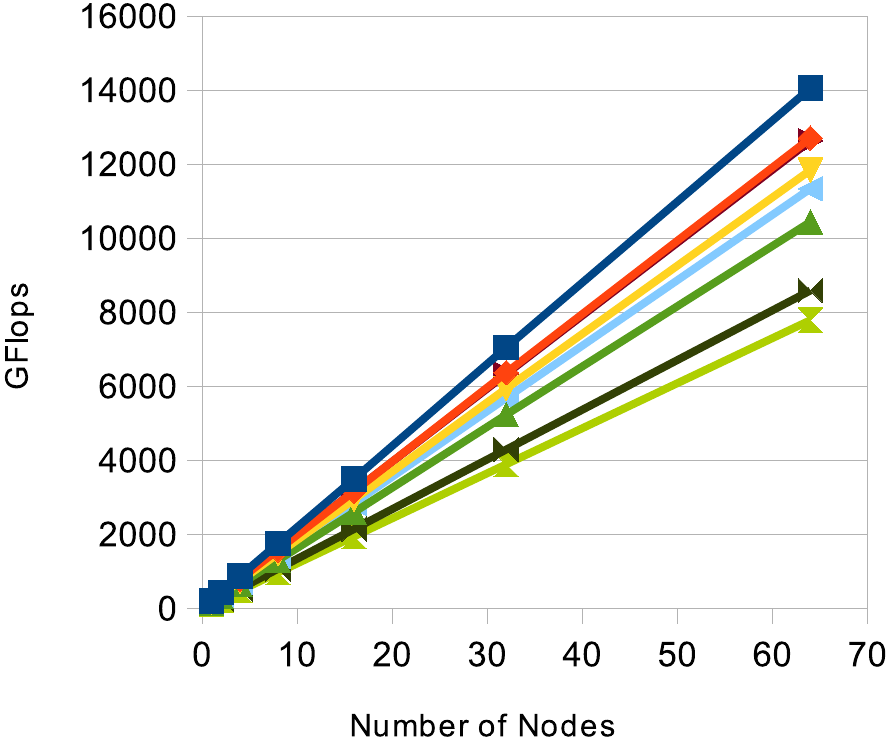}
\par\end{centering}}
\subfloat[Double precision.]{
\begin{centering}
\includegraphics[trim = 6mm 5mm 64mm 9mm, clip=true, height=5.5cm]{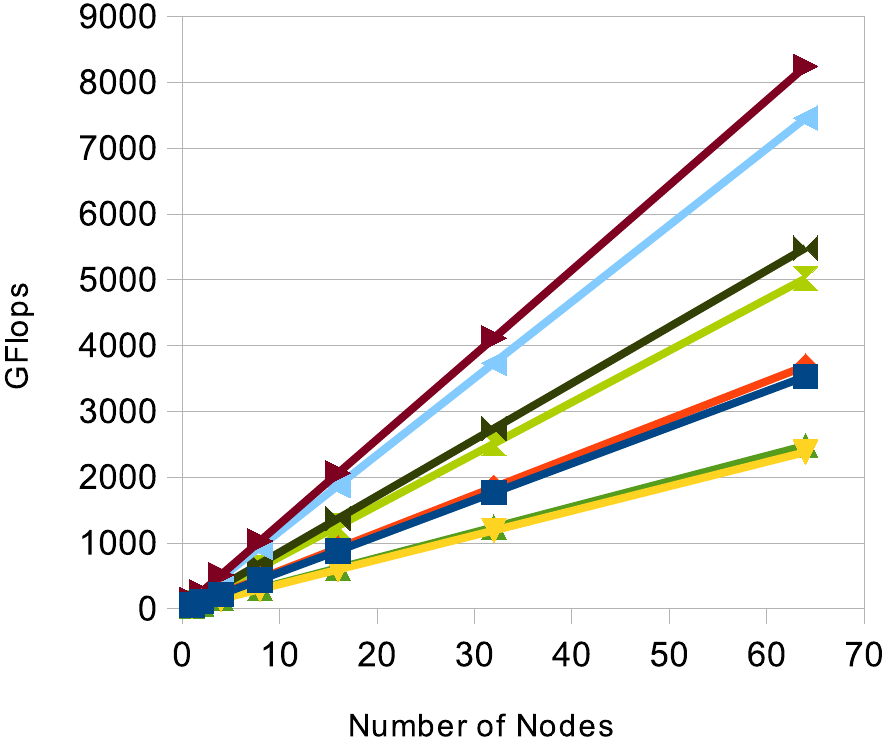}
\par\end{centering}}
\par\end{centering}
\caption{SU(3) weak Scaling for overrelaxation algorithm. Fixed node local lattice volume to $32^4$. Lattice volume: $32^3\times Nt$ with $ Nt = 32 \times (\text{Number of nodes})$. SP/DP means single/double precision, Tex means using Texture memory.}
\label{fig:overrelaxation_landau_weakscaling}
\end{figure}

\begin{figure}[h]
\begin{centering}
\subfloat{
\begin{centering}
\includegraphics[trim = 6mm 6mm 57mm 8mm, clip=true, height=5.2cm]{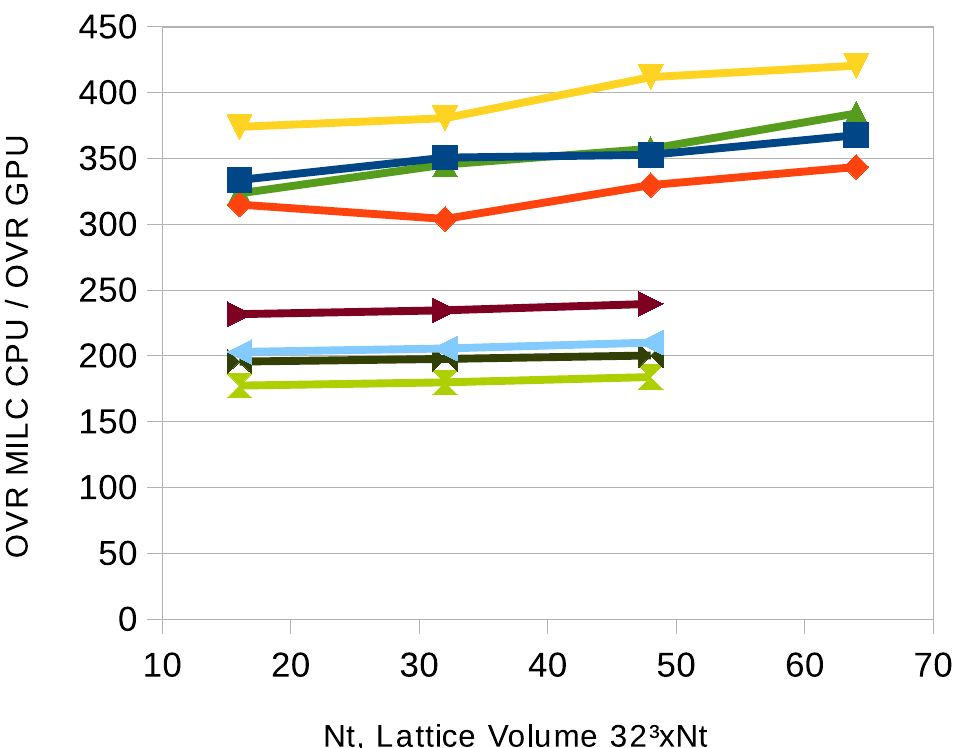}
\par\end{centering}}
\subfloat{
\begin{centering}
\quad\includegraphics[trim = 110mm 0mm 0mm 24mm, clip=true, height=3.9cm]{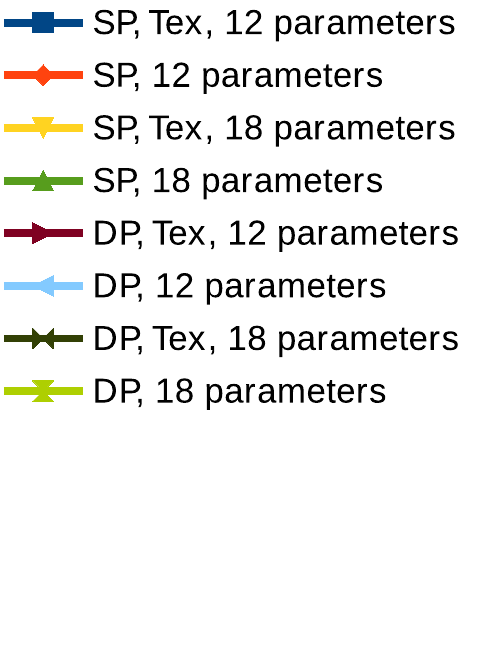}
\par\end{centering}}
\par\end{centering}
    \caption{Single GPU speed up over CPU MILC code in single node for the Landau gauge fixing with overrelaxation.  Using only CUDA atomic operations in single precision. SP/DP means single/double precision, Tex means using Texture memory.}
    \label{fig:GPU_speedUp_MILC}
\end{figure}

The performance results for the Landau gauge fixing using the overrelaxation algorithm are presented in 
\cref{fig:perf_overrelaxation_landau,fig:overrelaxation_landau_weakscaling} 
for single and multi-GPU.
The use of CUDA atomic operations, in single precision allow a speedup of $1.1-1.4\times$, although for double precision not using CUDA atomics we obtain a speedup of $2-2.4\times$, \cref{fig:overrelaxation_landau_weakscaling}. 
The double precision atomic additions is not yet supported natively by the hardware.
In \cref{fig:GPU_speedUp_MILC}, we compare the single GPU performance with the CPU MILC code, with and without taking in account the copies from and to the GPU and the gauge field reorder. 
Without taking into account the copies to and from GPU, the GPU performance is around 350/200 times faster than CPU MILC code in single/double precision.

In \cref{fig:overrelaxation_landau_weakscaling} we present the performance results for the weak scaling of the overrelaxation algorithm for a fixed local lattice volume of $32^4$  The results show a good performance scaling.

\begin{figure}[h]
\begin{centering}
\subfloat{
\includegraphics[trim = 110mm 56mm 0mm 24mm, clip=true, height=0.7cm]{figures/BW_coulomb_OVR_32_4_1000iter_gflops}
\includegraphics[trim = 110mm 46mm 0mm 34mm, clip=true, height=0.7cm]{figures/BW_coulomb_OVR_32_4_1000iter_gflops}
\includegraphics[trim = 110mm 34mm 0mm 46mm, clip=true, height=0.7cm]{figures/BW_coulomb_OVR_32_4_1000iter_gflops}
\includegraphics[trim = 110mm 24mm 0mm 56mm, clip=true, height=0.7cm]{figures/BW_coulomb_OVR_32_4_1000iter_gflops}}
\setcounter{subfigure}{0}

\vspace{-0.3cm}\subfloat[Single GPU performance.\label{subfig:coulomb_single_ovr}]{
\begin{centering}
\includegraphics[trim = 6mm 0mm 56mm 9mm, clip=true, height=5.5cm]{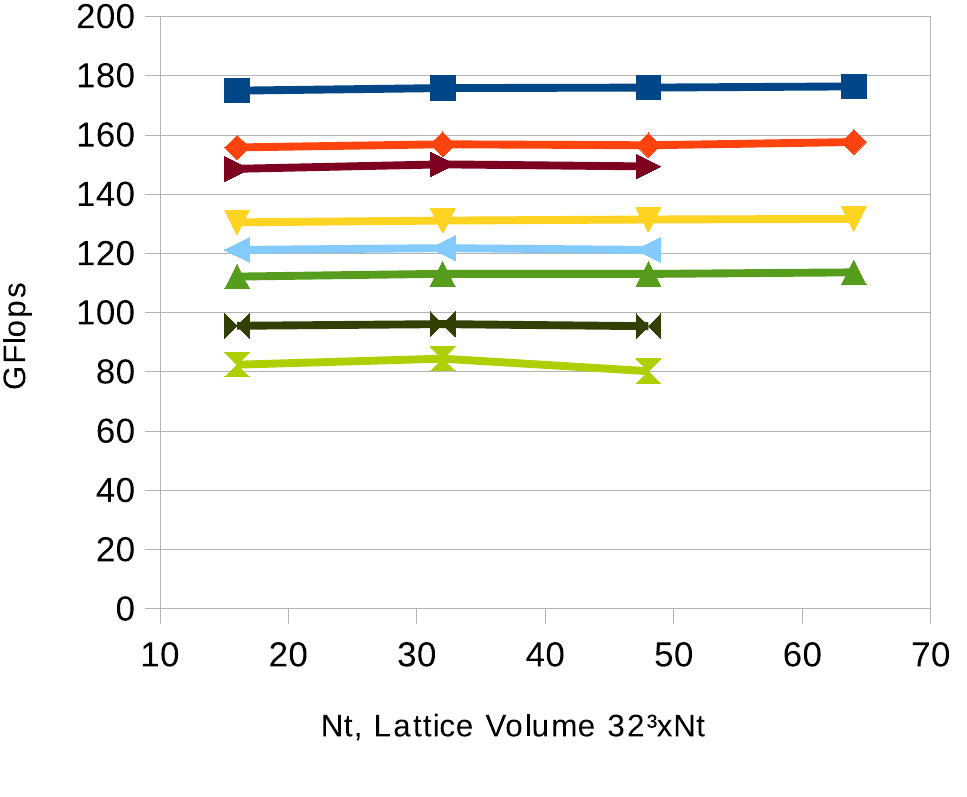}
\par\end{centering}}
\subfloat[Multi-GPU performance, $32^4$ lattice volume.\label{subfig:coulomb_multi_ovr}]{
\begin{centering}
\includegraphics[trim = 6mm 0mm 56mm 9mm, clip=true, height=5.5cm]{figures/BW_coulomb_OVR_32_4_1000iter_gflops}
\par\end{centering}}

\par\end{centering}
    \caption{SU(3) performance results for Coulomb gauge fixing with overrelaxation. Using only CUDA atomic operations in single precision. SP/DP means single/double precision, Tex means using Texture memory.}
    \label{fig:perf_coulomb_ovr}
\end{figure}

In \cref{fig:perf_coulomb_ovr} the performance results for the Coulomb gauge fixing with 
overrelaxation in single GPU are presented. Although we are expecting similar or better performance results for the Coulomb gauge fixing in comparison with the Landau gauge fixing, we obtained better performance for Landau gauge fixing in single precision. A detailed analysis of this behavior was not done yet.

The multi-GPU performance of gauge fixing with overrelaxation, \cref{subfig:landau_multi_ovr,subfig:coulomb_multi_ovr}, shows a good performance scaling up to four nodes for a lattice volume of $32^4$. 
For eight nodes, the performance shows a visible decrease due to the weak overlap in MPI communications and interior gauge links update. 
With eight nodes, MILC splits the lattice by half in $Y$, $Z$ and $T$ dimensions.

\begin{figure}[h]
\begin{centering}
\subfloat{
\includegraphics[trim = 110mm 56mm 0mm 24mm, clip=true, height=0.7cm]{figures/BW_coulomb_OVR_32_4_1000iter_gflops}
\includegraphics[trim = 110mm 46mm 0mm 34mm, clip=true, height=0.7cm]{figures/BW_coulomb_OVR_32_4_1000iter_gflops}
\includegraphics[trim = 110mm 34mm 0mm 46mm, clip=true, height=0.7cm]{figures/BW_coulomb_OVR_32_4_1000iter_gflops}
\includegraphics[trim = 110mm 24mm 0mm 56mm, clip=true, height=0.7cm]{figures/BW_coulomb_OVR_32_4_1000iter_gflops}}
\setcounter{subfigure}{0}

\vspace{-0.3cm}\subfloat[Landau gauge fixing.]{
\begin{centering}
\includegraphics[trim = 6mm 6mm 58mm 9mm, clip=true, height=5.5cm]{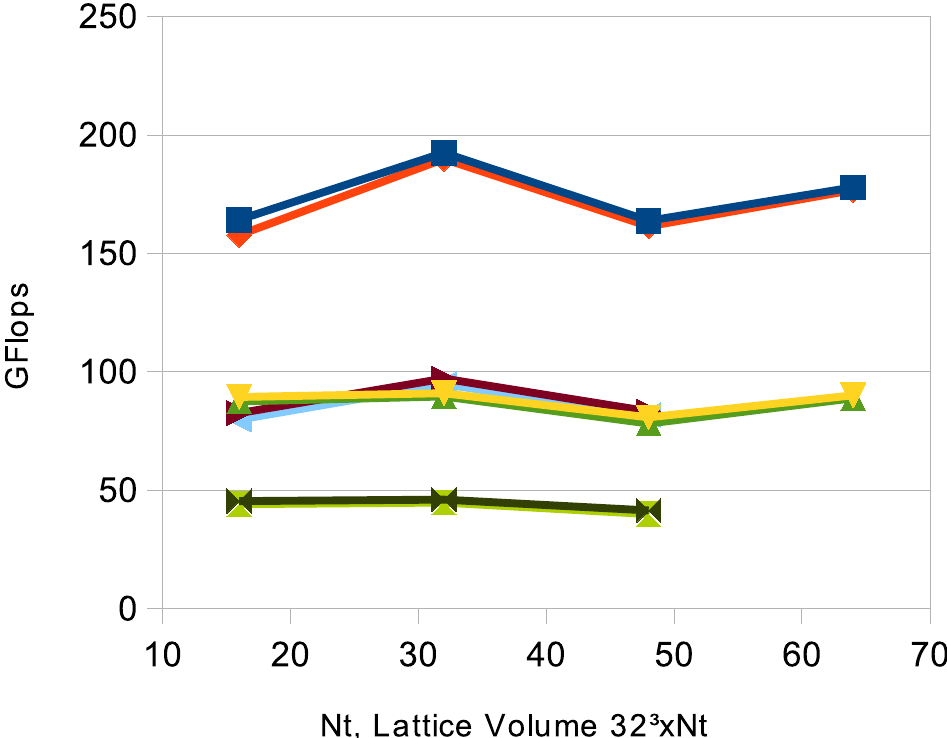}
\par\end{centering}}
\subfloat[Coulomb gauge fixing.]{
\begin{centering}
\includegraphics[trim = 6mm 6mm 58mm 9mm, clip=true, height=5.5cm]{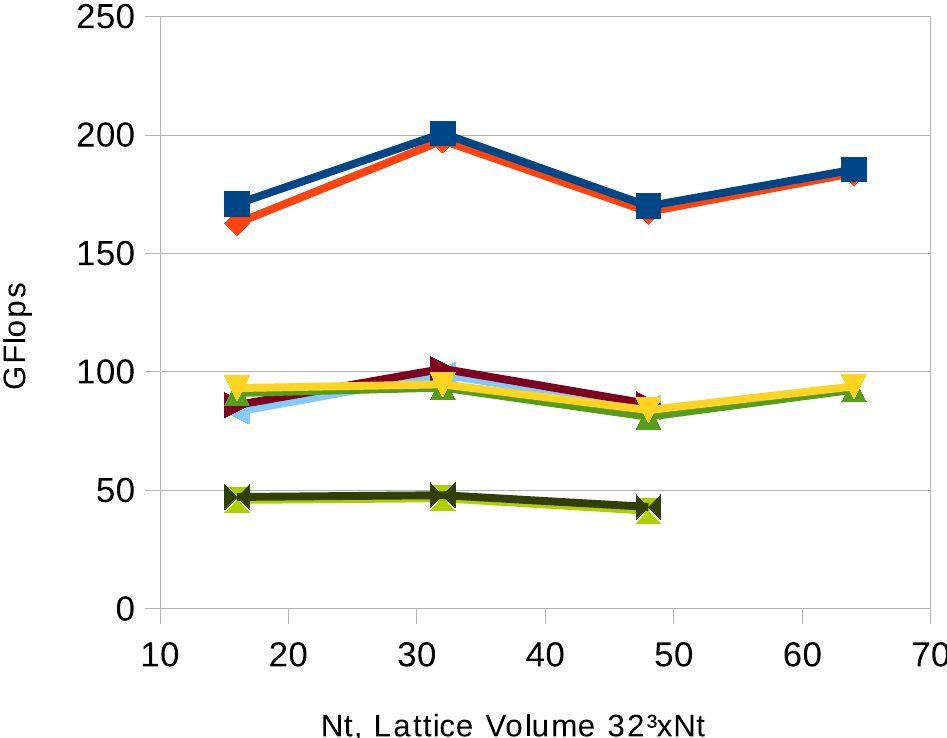}
\par\end{centering}}

\par\end{centering}
    \caption{SU(3) performance results for gauge fixing with steepest descent algorithm with Fourier acceleration in single GPU. SP/DP means single/double precision, Tex means using Texture memory.}
    \label{fig:BW_FFT_32_3_1000iter}
\end{figure}

The performance results for the steepest descent algorithm for Landau and Coulomb gauge fixing are presented in \cref{fig:BW_FFT_32_3_1000iter}. Further tests have to be done in order to better understand the performance fluctuations in changing the volume. These fluctuations are not visible when using the gauge fixing with overrelaxation, see \cref{subfig:landau_single_ovr,subfig:coulomb_single_ovr}.

\section{Conclusion}

The SU(N) code, with N>2, for Coulomb and Landau gauge fixing was implemented on CUDA.
The SU(3) code also supports 8 and 12 real number parametrization to store the gauge array in GPU memory.
Only the gauge fixing with overrelaxation method supports multi-GPUs using MPI.
In general, the steepest descent algorithm with Fourier acceleration converges faster than overrelaxation algorithm.
However, this method requires $1.5\times$ more memory than the overrelaxation.
This library is currently being ported to QUDA library, \cite{qudalibrary}.

\acknowledgments
This work is supported by the NSF award PHY-1212270. 
This research is part of the Blue Waters sustained-petascale computing project, which is supported by the National Science Foundation (awards OCI-0725070 and ACI-1238993) and the state of Illinois. Blue Waters is a joint effort of the University of Illinois at Urbana-Champaign and its National Center for Supercomputing Applications.
This work is also part of the Lattice QCD on Blue Waters PRAC allocation
supported by the National Science Foundation award ACI 0832315.



\begin{thebibliography}{99}
\expandafter\ifx\csname url\endcsname\relax
  \def\url#1{\texttt{#1}}\fi
\expandafter\ifx\csname urlprefix\endcsname\relax\def\urlprefix{URL }\fi
\expandafter\ifx\csname href\endcsname\relax
  \def\href#1#2{#2} \def\path#1{#1}\fi

\bibitem{Mandula:1990vs}
J.~E. Mandula, M.~Ogilvie, {Efficient gauge fixing via overrelaxation},
  Phys.Lett. B248 (1990) 156--158.

\bibitem{Davies:1987vs}
C.~Davies, G.~Batrouni, G.~Katz, A.~S. Kronfeld, G.~Lepage, et~al., {Fourier
  acceleration in lattice gauge theories. I. Landau gauge fixing}, Phys.Rev.
  D37 (1988) 1581.

\bibitem{Schrock:2012fj}
M.~Schröck, H.~Vogt, {Coulomb, Landau and Maximally Abelian Gauge Fixing in
  Lattice QCD with Multi-GPUs}, Comput.Phys.Commun. 184 (2013) 1907--1919.

\bibitem{Cardoso:2012pv}
N.~Cardoso, P.~J. Silva, P.~Bicudo, O.~Oliveira, {Landau Gauge Fixing on GPUs},
  Comput.Phys.Commun. 184 (2013) 124--129.

\bibitem{Cucchieri:1998ew}
A.~Cucchieri, T.~Mendes, {A Multigrid implementation of the Fourier
  acceleration method for Landau gauge fixing}, Phys.Rev. D57 (1998)
  3822--3826.
  
\bibitem{MILC}
{MIMD Lattice Computation (MILC)}
  \url{http://www.physics.indiana.edu/~sg/milc.html}.

\bibitem{qudalibrary}
Quda: A library for qcd on gpus, \url{http://lattice.github.io/quda/}.

\end{thebibliography}

\end{document}